\documentclass[twocolumn]{autart}    % Enable this line and disable the 
\usepackage{graphicx}          % Include this line if your 
\usepackage{url}
\usepackage{amsfonts}
\usepackage{amssymb}
\usepackage{amsmath}
\usepackage{subcaption}
\usepackage[round]{natbib}
\usepackage{color}

\newtheorem{ex}{Example}

\begin{document}

\begin{frontmatter}
%\runtitle{Insert a suggested running title}  % Running title for regular 
                                              % papers but only if the title  
                                              % is over 5 words. Running title 
                                              % is not shown in output.

\title{Symmetry reduction for dynamic programming\thanksref{footnoteinfo}} 

\thanks[footnoteinfo]{A preliminary version of this work was presented at the 2017 American Control Conference. Corresponding author J.~Maidens.}

\author[Ryerson]{John Maidens}\ead{maidens@ryerson.ca},   
\author[Safran]{Axel Barrau}\ead{axel.barrau@mines-paristech.fr},              
\author[Mines]{Silv\`{e}re Bonnabel}\ead{silvere.bonnabel@mines-paristech.fr}, and 
\author[Berkeley]{Murat Arcak}\ead{arcak@eecs.berkeley.edu}   

\address[Ryerson]{Department of Mechanical \& Industrial Engineering,
Ryerson University, 350 Victoria Street, Toronto, ON, M5B 2K3, Canada}                                              
\address[Safran]{SAFRAN TECH, Groupe Safran, Rue des Jeunes Bois - Ch\^{a}teaufort, 78772 Magny Les Hameaux, France}             
\address[Mines]{MINES ParisTech, PSL Research University, Centre for Robotics, 60 bd Saint-Michel, 75006 Paris, France}       
\address[Berkeley]{Department of Electrical Engineering \& Computer Sciences,
University of California, Berkeley, 569 Cory Hall, Berkeley, CA, 94720 USA} 
          

\begin{abstract}                          % Abstract of not more than 200 words.
We present a method of exploiting symmetries of discrete-time optimal control problems to reduce the dimensionality of dynamic programming iterations. The results are derived for systems with continuous state variables, and can be applied to systems with continuous or discrete symmetry groups. We prove that symmetries of the state update equation and stage costs induce corresponding symmetries of the optimal cost function and the optimal policies. We then provide a general framework for computing the optimal cost function based on gridding a space of lower dimension than the original state space. This method does not require algebraic manipulation of the state update equations; it only requires knowledge of the symmetries that the state update equations possess. 
Since the method can be performed without any knowledge of the state update map beyond being able to evaluate it and verify its symmetries, this enables the method to be applied in a wide range of application problems. We illustrate these results on two six-dimensional optimal control problems that are computationally difficult to solve by dynamic programming without symmetry reduction. 
\end{abstract}

\end{frontmatter}

\section{Introduction}
The dynamic programming algorithm for computing optimal control policies has, since its development, been known to suffer from the ``curse of dimensionality'' \citep{Bellman57}. Its applicability in practice is typically limited to systems with four or five continuous state variables because the number of points required to grid a space of $n$ continuous state variables increases exponentially with the state dimension $n$. This complexity has led to a collection of algorithms for approximate dynamic programming, which scale to systems with larger state dimension but lack the guarantees of global optimality of the solution associated with the original dynamic programming algorithm \citep{Bellman59, Bertsekas12, Powell07, Powell16}. 

In practice, many real-world systems exhibit symmetries that can be exploited to reduce the complexity of system models. Symmetry reduction has found applications in fields ranging from differential equations \citep{Clarksonz94, Bluman13} to model checking \citep{Emerson96, Kwiatkowska06}. In control engineering, symmetries have been exploited to improve control of mechanical systems \citep{Marsden90, Bloch96, Bullo99}, develop more reliable state estimators \citep{Barrau14}, study the controllability of multiagent systems \citep{Rahmani09} and to reduce the complexity of stability and performance certification for interconnected systems \citep{Arcak16, Rufino17}. Symmetry reduction has also been applied to the computation of optimal control policies for continuous-time systems in \citep{Grizzle84, Ohsawa13} and Markov decision processes (MDPs) in \citep{Zinkevich01, Narayanamurthy07}. 

In this paper, we present a theory of symmetry reduction for the optimal control of discrete-time, stochastic nonlinear systems with continuous state variables. This reduction allows dynamic programming to be performed in a lower-dimensional state space. Since the computational complexity of a dynamic programming iteration increases exponentially with state dimension, this reduction significantly decreases computational burden. Further, our proposed method does not rely on an explicit transformation of the state update equations, making the method applicable in situations where a such a transformation is difficult or impossible to find analytically. 

We present two theorems that summarize our method of symmetry reduction. Theorem \ref{thm:symmetric_cost_and_policy} describes how symmetries of the system dynamics imply symmetries of the optimal cost and optimal policy functions. Theorem \ref{thm:DP_reduced} then describes a method of computing the cost function based on reduced coordinate system that depends on fewer state variables. 

This paper builds on the work we presented in the conference paper \citep{Maidens17-ACC}. The most substantial improvement is the additional theoretical results presented in Sections \ref{sec:moving_frames} and \ref{sec:reduced_coordinates}. The conference version presented an \textit{ad hoc} symmetry reduction for a magnetic resonance imaging (MRI) application, but did not provide a general methodology for computing the coordinate reduction. This paper addresses this shortcoming by presenting a general method based on the moving frame formalism, which leads to the general symmetry reduction formula presented in Theorem \ref{thm:DP_reduced}. Additionally, the MRI example has been reworked to match this new formalism, and the numerical implementation and graphs of the numerical solution have been improved. We have also included two new extensions of this formalism to the case of equivariant costs in Section \ref{sec:equivariance} and to the synchronization problem of stochastic dynamic systems on matrix groups in Section \ref{sec:matrix_group}, along with examples to illustrate the algorithm in these contexts. 

This paper is organized as follows: in Section \ref{sec:dynamic_programming} we introduce notation and provide  background information both on dynamic programming for optimal control, and on the mathematical theory of symmetries. In Section \ref{Main}, we derive our main theoretical results, that is, we prove that control system symmetries induce symmetries of the optimal cost function and optimal control policy, and then leverage the result to present a general method of performing dynamic programming in reduced coordinates. In Section \ref{sec:vehicles_cooperative} we apply the algorithm to a cooperative control problem for two Dubins vehicles using a Lie group formulation. In Section \ref{sec:MRI} we apply symmetry reduction to compute the solution of an optimal control problem arising in dynamic MRI acquisition. Code to reproduce the computational results in this paper is available at \url{https://github.com/maidens/Automatica-2017}.

\section{Dynamic Programming and Symmetries}
\label{sec:dynamic_programming}
In this section, we first recall the main features of dynamic programming for optimal control of stochastic discrete time systems. Then we introduce our problem and provide the reader with a primer on the classical theory of symmetries. We also introduce the notion of invariant control systems with invariant costs. 

\subsection{Dynamic programming for optimal control of stochastic systems}
We begin by introducing dynamic programming for finite horizon optimal control following the notation of \citep{Bertsekas05}. We consider a discrete-time dynamical system
\begin{align}
	x_{k+1} = f_k(x_k, u_k, w_k), \quad k = 0, 1, \dots, N-1\label{syst_initial:eq}
\end{align}
where $x_k \in \mathcal{X} \subseteq \mathbb{R}^n$ is the system state, $u_k \in \mathcal{U} \subseteq \mathbb{R}^m$ is the control variable to be chosen at time $k$, $w_k \in \mathcal{W} \subseteq \mathbb{R}^\ell$ are independent continuous random variables each with density $p_k$, and $N \in \mathbb{Z}_+$ is a finite control horizon. Associated with this system is an additive cost function 
\[
    g_N(x_N) + \sum_{k=0}^{N-1} g_k(x_k, u_k, w_k) 
\]
that we wish to minimize through our choice of $u_k$. We define a \emph{control system} to be a tuple $\mathcal{S} = (\mathcal{X}, \mathcal{U}, \mathcal{W}, p, f, g, N)$ where $p = \prod_{k=0}^{N-1} p_k$ is the joint density of the random variables $w_k$. 

We consider a class of control policies $\pi = \{ \mu_0, \dots, \mu_{N-1} \}$ where $\mu_k: \mathcal{X} \to \mathcal{U}$ maps observed states to admissible control inputs. Given an initial state $x_0$ and a control policy $\pi$, we define the expected cost under this policy as
\[
  J_\pi(x_0) = \mathbb{E}\left[g_N(x_N) + \sum_{k=0}^{N-1} g_k(x_k, \mu_k(x_k), w_k)\right].  
\]
An optimal policy $\pi^*$ is defined as one that minimizes the expected cost:
\[
  J_{\pi^*}(x_0) = \min_{\pi \in \Pi} J_\pi(x_0) 
\]
where $\Pi$ denotes the set of all admissible control policies. The optimal cost function, denoted $J^*(x_0)$, is defined to be the expected cost corresponding to an optimal policy. 

As in \citep{Bertsekas05}, we use $\min$ to denote the infimum value regardless of whether there is a policy $\pi \in \Pi$ that achieves a minimum. In the example problems presented in Sections \ref{sec:vehicles_cooperative} and \ref{sec:MRI}, the existence of a minimum is guaranteed by compactness or finiteness arguments respectively. In the general case, the optimal cost function $J^*$ can be computed using the dynamic programming algorithm regardless of the existence of minimizers, but the existence of an optimal policy $\pi^*$ requires that a minimum be achieved for each $x_k \in \mathcal{X}$. 

We quote the following result due to Bellman from \citep{Bertsekas05}: 

\begin{prop}[Dynamic Programming]
For every initial state $x_0$, the optimal cost $J^*(x_0)$ is equal to $J_0(x_0)$, given by the last step of the following algorithm, which proceeds backward in time from period $N-1$ to period 0:
\begin{equation}
\resizebox{.9\hsize}{!}{$
\begin{split}
J_N(x_N) &= g_N(x_N) \\
J_k(x_k) &= \min_{u_k \in \mathcal{U}} \mathbb{E} \Bigg[ g_k(x_k, u_k, w_k) + J_{k+1}\Big(f_k(x_k, u_k, w_k)\Big)\Bigg] \\
&  \quad \quad k=0, 1, \dots, N-1, 
\end{split}
$}
\label{eq:DP}
\end{equation}
where the expectation is taken with respect to the probability distribution of $w_k$ defined by the density $p$. Furthermore, if there exists $u_k^*$ minimizing the right hand side of \eqref{eq:DP} for each $x_k$ and $k$, then the policy $\pi^* = \{\mu_0^*, \dots, \mu_{N-1}^*\}$ where $\mu_k^*(x_k) = u_k^*$ is optimal. 
\end{prop}

The intermediate functions $J_k(x_k)$ for $k > 0$ computed in this manner represent the optimal cost of the tail subproblem beginning at $x_k$. The optimal cost of the entire problem is given by the function $J^*(x_0) = J_0(x_0)$ obtained when this recursion terminates.

\subsection{Invariant system with invariant costs} 
\label{sec:symmetry_reduction} 

We first recall the definition of a transformation group for a control system,  as in \citep{Martin04, Jakubczyk98, Respondek02}. See  \cite{olver1999classical} for the more general theory. 

\begin{defn}[Transformation group]
A transformation group on $\mathcal{X} \times \mathcal{U} \times \mathcal{W}$ is set of tuples $h_\alpha = (\phi_\alpha, \chi_\alpha, \psi_\alpha)$ parametrized by elements $\alpha$ of a Lie group $\mathcal{G}$  having dimension $r$, such that the functions $\phi_\alpha: \mathcal{X} \to \mathcal{X}$, $\chi_\alpha: \mathcal{U} \to \mathcal{U}$ and $\psi_\alpha: \mathcal{W} \to \mathcal{W}$ are all $C^1$ diffeomorhpisms and satisfy:
\begin{itemize}
\item $\phi_e(x) = x$, $\chi_e(u) = u$, $\psi_e(w) = w$ when $e$ is the identity of the group $\mathcal{G}$ and
\item $\phi_{a * b}(x) = \phi_a \circ \phi_b (x)$, $\chi_{a * b}(u) = \chi_a \circ \chi_b (u)$, $\psi_{a * b}(x) = \psi_a \circ \psi_b (x)$ for all $a, b \in \mathcal{G}$ where $*$ denotes the group operation and $\circ$ denotes function composition. 
\end{itemize} 
\end{defn}
To simplify notation we will sometimes suppress the subscripts $\alpha$.
In the present paper, we will consider the following class of systems and cost functions. 
\begin{defn}
{\bf (Invariant control system with invariant costs)}
A control system $\mathcal{S}$ is $\mathcal{G}$-invariant with $\mathcal{G}$-invariant costs if for all $\alpha \in \mathcal{G}$, $x_k \in \mathcal{X}$, $u_k \in \mathcal{U}$ and $w_k \in \mathcal{W}$ we have: 
\[
\begin{split}
  \phi^{-1} \circ f_k( \phi(x_k), \chi(u_k), \psi(w_k)) &= f_k(x_k, u_k, w_k), \\
  & \quad k = 0, 1, \dots, N-1\\
  g_k(\phi(x_k), \chi(u_k), \psi(w_k)) &= g_k(x_k, u_k, w_k),  \\
  & \quad k = 0, 1, \dots, N-1, \\
  g_N(\phi(x_N)) &= g_N(x_N), \text{ and } \\
  p_k(\psi(w_k)) |\det D\psi(w_k)| &= p_k(w_k) \\
  & \quad k = 0, 1, \dots, N-1 
\end{split}
\]
where $D\psi$ denotes the Jacobian of $\psi$. 
\end{defn}
The rationale is simple: For any fixed $\alpha\in\mathcal{G}$, consider the change of variables $X_k=\phi_\alpha(x_k)$, $U_k=\chi_\alpha(u_k) $, $W_k=\psi_\alpha(w_k)$. Then, we have 
$$
X_{k+1} = f_k(X_k, U_k, W_k), \quad k = 0, 1, \dots, N-1,
$$and for $ k = 0, 1, \dots, N-1$ we have also $g_k(X_k, U_k, W_k)=g_k(x_k,u_k,w_k)$. As a result, if $u_1,\dots,u_{N-1}$ is a series of controls that minimize $J(x_0)$, then one can expect $U_1,\dots,U_{N-1}$ to minimize $J(X_0)$, under some assumptions on the noise. As a result, the optimal control problem needs only be solved once for all initial conditions belonging to the set $\{\phi_\alpha(x_0)|\alpha\in\mathcal G\}$, reducing the initial $n$ dimensional problem to a $n-r$ dimensional problem. The present paper derives a theory for such symmetry reduction in dynamic programming, and provides various examples of engineering interest.

\subsection{Cartan's moving frame method} 
\label{sec:moving_frames}

To find a reduced coordinate system in which to perform dynamic programming, we will use the moving frame method of Cartan \citep{Cartan37}. In general, this method only results in a local coordinate transformation as it relies on the implicit function theorem. In this paper we will focus on a transformation within a single coordinate chart. For many practical problems, including both examples in this paper, the transformation computed using this method extends to all of $\mathcal{X}$ with the exception of a lower-dimensional submanifold of the state space. In such cases, only a single coordinate chart is required for the purpose of gridding the entire state space for dynamic programming. 

We briefly introduce the moving frame method following the presentation in \citep{Bonnabel08}. Consider an $r$-dimensional transformation group (with $r \le n$) acting on $\mathcal{X}$ via the diffeomorphisms $(\phi_\alpha)_{\alpha \in \mathcal{G}}$. Assume we can split $\phi_\alpha$ as $(\phi_\alpha^a, \phi_\alpha^b)$ with $r$ and $n-r$ components respectively so that $\phi_\alpha^a$ is an invertible map.   Then, for some $c$ in the range of $\phi^a$, we define a coordinate cross section to the orbits $\mathcal{C} = \{x: \phi_e^a(x) = c\}$. This cross section is an $n-r$-dimensional submanifold of $\mathcal{X}$. Assume moreover that for any point  $x \in \mathcal{X}$, there is a unique group element $\alpha \in \mathcal{G}$ such that $\phi_\alpha(x) \in \mathcal{C}$. Such $\alpha$ will be denoted $\gamma(x)$, and the map $\gamma: \mathcal{X} \to \mathcal{G}$ will be called moving frame.

A moving frame can be computed by solving the normalization equation:
\[
\phi_{\gamma(x)}^a(x) = c. 
\]
Define the following map $\rho: \mathcal{X} \to \mathbb{R}^{n-r}$ as 
\[
\rho(x) = \phi^b_{\gamma(x)}(x). 
\]
Note that, for all $\alpha\in\mathcal G$ we have $\rho(\phi_\alpha(x))=\rho(x)$, that is, the components of $\rho$ are $\emph{invariant}$ to the group action on the state space. Further, due to our assumptions, the restriction of $\rho$ to $\mathcal{C}$ is injective. We denote this restricted function $\bar \rho$, and it will serve as a reduced coordinate system to solve the invariant optimal control problem.

\section{Main Results}\label{Main}

In order to combat the ``curse of dimensionality'' associated with performing dynamic programming in high-dimensional systems, we describe a method to reduce the system's dimension by exploiting symmetries in the dynamics and stage costs. 

\subsection{Symmetries imply equivalence classes of optimal policies}

\begin{thm}
{\bf (Symmetries of the optimal cost and policy)}
\label{thm:symmetric_cost_and_policy}
Let $\mathcal{G}$ be a group and let $\mathcal{S}$ be a $\mathcal{G}$-invariant control system with $\mathcal{G}$-invariant costs. Then the optimal cost functions $J_k(x_0)$ satisfy the symmetry relations
\[
  J_k  = J_k \circ \phi_\alpha 
\]
for any $k = 0, \dots, N$ and any $\alpha \in \mathcal{G}$. Furthermore, if $\pi^* = \{\mu_0^*, \dots, \mu_{N-1}^*\}$ is an optimal policy then so is $\tilde \pi^* := \{ \chi_\alpha \circ \mu_0^* \circ \phi^{-1}_\alpha, \dots,  \chi_\alpha \circ \mu_{N-1}^* \circ \phi^{-1}_\alpha \}$ for any $\alpha \in \mathcal{G}$. 
\end{thm} 

\begin{pf} We prove this by induction on $k$ proceeding backward from the base case $k=N$. 
First, note that 
\[
J_N(x_N) = g_N(x_N) = g_N(\phi(x_N)) = J_N(\phi(x_N)). 
\]
Now, suppose that for some $k \in \{0, \dots, N-1\}$ we have $J_{k+1}(x_{k+1}) = J_{k+1}(\phi(x_{k+1}))$ for all $x_{k+1} \in \mathcal{X}$. Then for any $x_k \in X$, and $u_k \in \mathcal{U}$ we have
\[
\resizebox{1.05\hsize}{!}{$
\begin{split}
& \mathbb{E} \Bigg[ g_k(x_k, u_k, w_k) + J_{k+1}(f_k(x_k, u_k, w_k)) \Bigg] \\
&= \int_\mathcal{W} \Bigg[ g_k(x_k, u_k, w_k) + J_{k+1}(f_k(x_k, u_k, w_k)) \Bigg] p_k(w_k) dw_k \\
&= \int_\mathcal{W} \Bigg[ g_k(\phi(x_k), \chi(u_k), \psi(w_k)) + J_{k+1}(\phi^{-1} \circ f_k(\phi(x_k), \chi(u_k), \psi(w_k))) \Bigg] p_k(w_k) dw_k \\
&= \int_\mathcal{W} \Bigg[ g_k(\phi(x_k), \chi(u_k), \psi(w_k)) + J_{k+1}(f_k(\phi(x_k), \chi(u_k), \psi(w_k))) \Bigg] p_k(w_k) dw_k \\
&= \int_\mathcal{W} \Bigg[ g_k(\phi(x_k), \chi(u_k), \psi(w_k)) + J_{k+1}(f_k(\phi(x_k), \chi(u_k), \psi(w_k))) \Bigg] p_k(\psi(w_k)) |\det D\psi (x_k)| dw_k \\
\end{split}
$}
\]
\[
\resizebox{\hsize}{!}{$
\begin{split}
&= \int_{\psi(\mathcal{W})} \Bigg[ g_k(\phi(x_k), \chi(u_k), \tilde w_k) + J_{k+1}(f_k(\phi(x_k), \chi(u_k), \tilde w_k)) \Bigg] p_k(\tilde w_k) d\tilde w_k \\
&= \int_\mathcal{W} \Bigg[ g_k(\phi(x_k), \chi(u_k), w_k) + J_{k+1}(f_k(\phi(x_k), \chi(u_k), w_k)) \Bigg] p_k(w_k) dw_k \\
&= \mathbb{E} \Bigg[ g_k(\phi(x_k), \chi(u_k), w_k) + J_{k+1}(f_k(\phi(x_k), \chi(u_k), w_k)) \Bigg]
\end{split}
$} 
\]
where the change of variables $\tilde w_k$ is defined via $\tilde w_k = \psi(w_k)$ and the tildes are subsequently dropped. Therefore, from the one-step dynamic programming principle we have 
\[
\resizebox{\hsize}{!}{$
\begin{split}
J_k(x_k) &= \min_{u_k \in \mathcal{U}} \mathbb{E} \Bigg[ g_k(x_k, u_k, w_k) + J_{k+1}(f_k(x_k, u_k, w_k)) \Bigg] \\
&= \min_{u_k \in \mathcal{U}} \mathbb{E} \Bigg[ g_k(\phi(x_k), \chi(u_k), w_k) + J_{k+1}(f_k(\phi(x_k), \chi(u_k), w_k)) \Bigg] \\
&= \min_{\tilde u_k \in \chi(\mathcal{U})} \mathbb{E} \Bigg[ g_k(\phi(x_k), \tilde u_k, w_k) + J_{k+1}(f_k(\phi(x_k), \tilde u_k, w_k)) \Bigg] \\
&= J_k(\phi(x_k)). 
\end{split}
$}
\] 
Thus $J^* = J^* \circ \phi$. Now, if $\pi^* = \{\mu_0^*, \dots, \mu_{N-1}^*\}$ is an optimal policy and we denote $\tilde x_k = \phi(x_k)$ then for any $k \in \{0, \dots, N-1\}$ we have 
\[
\resizebox{\hsize}{!}{$
\begin{split}
J_k(\tilde x_k) &= J_k(x_k) \\
&= \mathbb{E}\Bigg[ g(x_k, \mu_k^*(x_k), w_k) + J_{k+1}(f_k(x_k, \mu_k^*(x_k), w_k)) \Bigg] \\
&= \mathbb{E} \Bigg[ g_k(\phi(x_k), \chi(\mu_k^*(x_k)), w_k) + J_{k+1}(f_k(\phi(x_k), \chi(\mu_k^*(x_k)), w_k)) \Bigg] \\
&= \mathbb{E} \Bigg[ g_k(\phi(x_k), \chi \circ \mu_k^* \circ \phi^{-1}(\phi(x_k)), w_k) + J_{k+1}(f_k(\phi(x_k), \chi \circ \mu_k^* \circ \phi^{-1} (\phi(x_k)), w_k)) \Bigg] \\
&= \mathbb{E} \Bigg[ g_k(\tilde x_k, \chi \circ \mu_k^* \circ \phi^{-1}(\tilde x_k), w_k) + J_{k+1}(f_k(\tilde x_k, \chi \circ \mu_k^* \circ \phi^{-1} (\tilde x_k), w_k)) \Bigg].
\end{split}
$}
\]
Thus $\tilde \pi^* := \{ \chi \circ \mu_0^* \circ \phi^{-1}, \dots,  \chi \circ \mu_{N-1}^* \circ \phi^{-1} \}$ is an optimal policy. \qed
\end{pf}

\subsection{Dynamic programming can be performed using reduced coordinates} 
\label{sec:reduced_coordinates}

Theorem \ref{thm:symmetric_cost_and_policy} readily implies the problem can be reduced, as all states along an orbit of $\mathcal{G}$ are equivalent in terms of cost, and that there are equivalence classes of optimal policies. So it suffices to only consider the cost corresponding to a single representative of each equivalence class, and to find a single representative of the optimal policy within each class. This can now easily be done using the injective map $\bar\rho: \mathcal{C}\to\mathbb{R}^{n-r}$.

For $\bar x \in \bar\rho(\mathcal{C})\subset \mathbb{R}^{n-r}$, let $z\in\mathcal C$ be such that $\bar x=\bar\rho(z)$, and define
\[
\bar J_k(\bar x) = J_k(z). 
\]
The following result  shows that the functions $J_k$ on the $n$-dimensional space $\mathcal{X} \subseteq \mathbb{R}^n$ are completely determined by the values of $\bar J_k$ on the subset $ \bar\rho(\mathcal{C})$ of $\mathbb{R}^{n-r}$. 

\begin{cor}
For any $x \in \mathcal{X}$ and $k = 0, \dots, N$, the cost function $J_k$ for the full problem can be computed in terms of the lower-dimensional cost function $\bar J_k$ as
\[
J_k(x) = J_k(\phi_{\gamma(x)}(x))=\bar J_k(\bar x), 
\]where $\bar x:=\bar\rho(\phi_{\gamma(x)}(x))$ is well defined as $\phi_{\gamma(x)}(x)\in\mathcal C$. 
\end{cor}
It is thus sufficient to have evaluated $\bar J$ at all points of $\bar\rho(\mathcal{C})\subset \mathbb{R}^{n-r}$ to be able to instantly evaluate $J$ at any point of $\mathcal X$. An optimal policy $\bar \pi^* = \{\bar \mu_k^*: \bar\rho(\mathcal{C}) \to \mathcal{U}\}_{k=0}^{N-1}$ can also be lifted to an optimal policy on the original state space via this method:
\[
\mu_k^*(x) = \bar \mu_k^*(\bar \rho(\phi_{\gamma(x)}(x))). 
\]
This allows optimal trajectories of the system to be computed in the original coordinates $\mathcal{X}$ via the lifted policy $\pi^* = \{\mu_0^*, \dots, \mu_{N-1}^*\}$. 

\begin{thm}{\bf (Dynamic programming in reduced coordinates)}
\label{thm:DP_reduced}
The reduced coordinates are in one to one correspondance with the cross-section $\mathcal C$. For any $\bar x \in \bar \rho(\mathcal{C})$, let $z\in\mathcal C$ satistisfy $\bar\rho(z)=\bar x$. Then in the reduced coordinates, the sequence $\bar J_k$ can be computed recursively via
\[
\resizebox{\columnwidth}{!}{$
\bar J_k(\bar x) = \min_{u_k \in \mathcal{U}} \mathbb{E}\left[ g_k(z, u_k, w_k) + \bar J_{k+1}(\rho(f_k(z, u_k, w_k)))\right]. 
$}
\]
\end{thm}
\begin{pf}
We have
\[
\resizebox{\columnwidth}{!}{$
\begin{split}
\bar J_k(\bar x) = J_k(z) 
&= \min_{u_k \in \mathcal{U}} \mathbb{E} \Bigg[ g_k(z, u_k, w_k) + J_{k+1}\Big(f_k(z, u_k, w_k)\Big)\Bigg] \\
&= \min_{u_k \in \mathcal{U}} \mathbb{E} \Bigg[ g_k(z, u_k, w_k) + \bar J_{k+1}\Big( \rho \circ f_k(z, u_k, w_k)\Big)\Bigg]. \qed
\end{split}
$}
\]
\end{pf}

\subsection{Case of equivariant costs}
\label{sec:equivariance}

So far, we have considered the costs to be invariant under transformation. We now briefly discuss how this can be generalized. The cost $g_k$ is said to be equivariant if there exists a family of diffeomorphisms $\varphi_\alpha$ such that  $g_k(\phi_\alpha(x_k), \chi_\alpha(u_k), \psi_\alpha(w_k)) = \varphi_\alpha\circ g_k(x_k, u_k, w_k)$. As we want the cost function $J$ to be equivariant too, we will need $\varphi_\alpha(\cdot)$ to be linear. Thus we will simply assume that $\varphi_\alpha$ is of the form $\varphi_\alpha(J)=l(\alpha)J$, that is, it is a scaling of the cost, where $l:\mathbb R_{>0}\to\mathbb R_{>0}$. For simplicity's sake, we consider here the problem to be noise free. Along the lines of the preceding sections it is easily proved that
$$
J_k(\phi(x))=\varphi\circ J_k(x)
$$
as already noticed in \citep{alvarez1998dynamic} for the case of homogeneous costs. Symmetry reduction can then be applied. We now give two tutorial examples.

\begin{ex}Consider the linear system 
$$
x_{x+1}=Ax_k+Bu_k
$$
with quadratic costs $g_k=x_k^TQx_k+u_k^TRu_k$. The system is invariant to scalings,  $\phi_\alpha(x)=\alpha x$, $\chi_\alpha(u)=\alpha u$, and the cost is equivariant letting $\varphi_\alpha(J)=\alpha J$, where $\alpha\in\mathcal G=\mathbb R_{>0}$. The unit sphere is a cross section to the orbits, and the normalization equation yields $\gamma(x)={1}/{||x||}$. Applying the results above, we see that the  controls  that minimize $J(x_0)$ are  $||x_0||u_1^*,\cdots,||x_0||u_{N-1}^*$, where $u_1^*,\cdots,u_{N-1}^*$ are those minimizing $J(\frac{x_0}{||x_0||})$. This agrees with the well known fact that the  optimal controller  for the problem above is the linear quadratic controller, and is indeed of the linear form $u_k=-F_kx_k$. 
\end{ex}

\begin{ex}Consider the following system and costs
$$
x_{x+1}=Ax_k+Bu_k,\quad g_k=h(x_k)+||u_k||_{1}
$$
where $||u_k||_{1}$ denotes the $L^1$ norm of $u_k$ and $h$ is a map satisfying $h(ax) = ah(x)$ for $a > 0$. Such costs may arise when one tries to force some controls to zero to create sparsity, a method known as $L^1$ regularization. This problem is challenging, particularly for nonconvex $h$. But according to the theory above, it is sufficient to solve it numerically for initial conditions lying on the unit sphere of the state space. 
\end{ex}

\subsection{Optimal formation control on Lie groups} 
\label{sec:matrix_group}

We now apply the theory presented in Section \ref{sec:symmetry_reduction} where the state space  $\mathcal{X} \subseteq \mathbb{R}^n$ is the Cartesian product of matrix Lie groups. Note that, straightforward modifications arise along the way as the state space and noise space are not vector spaces as in the theory above. The methodology is then applied to the synchronization of two non-holonomic cars in the presence of uncertainties.  

We model the system as a collection of $K$ agents, where the state of each agent evolves on a $r$-dimensional matrix Lie group $\mathcal{G}$. We assume that the evolution of the state of agent $j$ proceeds according to the equation 
\begin{equation}
X_{k+1}^j = X_k^j M(u_k^j) W_k^j 
\label{eq:Lie_group_dynamics}
\end{equation}
where $X_k$, $M(u_k)$, $W_k$ are all square matrices belonging to $\mathcal{G}$, $u_k$ is a control that lives in some finite dimensional vector space, and $W_k$ is the noise. The control objective is to reach a desired configuration, that is, a desired value for the relative  configurations of the agents $(X^1)^{-1} X^2,\dots,(X^{K-1})^{-1} X^K$, see e.g.,  \cite{sarlette2010coordinated} for more information.

Systems of this form are naturally invariant to left multiplication of all $X^j$ by some matrix $A \in \mathcal{G}$: 
\[
\phi_A(X) = \begin{bmatrix}AX^1 \\ \vdots \\ AX^K \end{bmatrix}
\]
where $X = (X^1, \dots, X^k) \in \mathcal{G}^K$. Letting $\chi(u^1,\dots,u^K)\equiv (u^1,\dots,u^K)$,  $\psi(W^1,\dots,W^K)\equiv (W^1,\dots,W^K)$, and the costs be of the form $\tilde g((X^1)^{-1} X^2,\dots,(X^{K-1})^{-1} X^K)+h(u^1,\dots,u^K)$, we get an invariant system with invariant costs.

One can define a cross section to the orbits by letting the first agent coordinates be equal to the identity matrix, that is, $\mathcal{C} = \{X \in \mathcal{G}^K: X^1 = I\}$. The normalization equation is given by $I = \phi^a_{\gamma(X)}(X) = \gamma(X) X^1$, hence the moving frame is given by $\gamma(X) = (X^1)^{-1}$. The invariants are computed as 
\[
\rho(X) = \phi^b_{\gamma(X)} = \begin{bmatrix} (X^1)^{-1} X^2 \\ \vdots \\ (X^1)^{-1} X^K \end{bmatrix}.
\]The optimal stochastic control problem can  then be solved in the reduced coordinate system defined by $\rho$,  reducing the state space from dimension $Kr$ to $(K-1)r$.

\section{Application I: Cooperative formation control for two stochastic Dubins vehicles}
\label{sec:vehicles_cooperative}

We consider two identical Dubins vehicles each with dynamics 
\[
\begin{split}
z_{k+1} &= z_k + v_k \cos \theta_k \\
y_{k+1} &= y_k + v_k \sin \theta_k \\
\theta_{k+1} &= \theta_k + \frac{1}{L} v_k \tan s_k + w_k
\end{split} 
\]
where $y_k$ and $z_k$ denote the two-dimensional position of the vehicle, $\theta_k$ denotes the heading of the vehicle, $v_k$ is a velocity input, $s_k$ is a steering angle input, and $w_k$ is independent, identically-distributed zero-mean Gaussian noise with variance $\sigma^2$, and $L$ is a parameter that determines the vehicle's steering radius. 

These dynamics can be embedded in the three-dimensional special Euclidean matrix Lie group $\mathcal{G} = SE(2)$, by defining the state 
\[
X_k = \begin{bmatrix} \cos \theta_k & -\sin \theta_k & z_k \\ \sin \theta_k & \cos \theta_k & y_k \\ 0 & 0 & 1 \end{bmatrix},
\]
input matrix
\[
M(v_k, s_k) = \begin{bmatrix} \cos(\frac{1}{L} v_k \tan s_k) & -\sin(\frac{1}{L} v_k \tan s_k) & v_k \\
\sin(\frac{1}{L} v_k \tan s_k) & \cos (\frac{1}{L} v_k \tan s_k) & 0\\ 0 & 0 & 1 \end{bmatrix},
\]
and noise matrix
\[
W_k = \begin{bmatrix} \cos w_k & -\sin w_k & 0 \\ \sin w_k & \cos w_k & 0 \\ 0 & 0 & 1 \end{bmatrix}, 
\]
with state update equation of the form \eqref{eq:Lie_group_dynamics}.

We wish to compute a control policy for a two-vehicle system, with states $X^1$ and $X^2$, where the controls can only take a finite number of values, and with terminal cost 
\[
J(X_0^1,X_0^2)=\mathbb E  \Bigg[g_N\bigl((X^1_N)^{-1} X^2_N\bigr)   \Bigg]
\]
where  $
g_N(X) = \arccos(X_{11})^2 + |\sqrt{X_{13}^2 + X_{23}^2} - 1 |$, that is, we want the vehicles to have the same heading, and follow each other at unit distance. Thanks to the theory developed above, the stochastic control problem is reduced from problem with a six dimensional state space to a problem with a three dimensional  state space only. 

\begin{figure}[ht!]
\begin{center}
\includegraphics[width=\columnwidth]{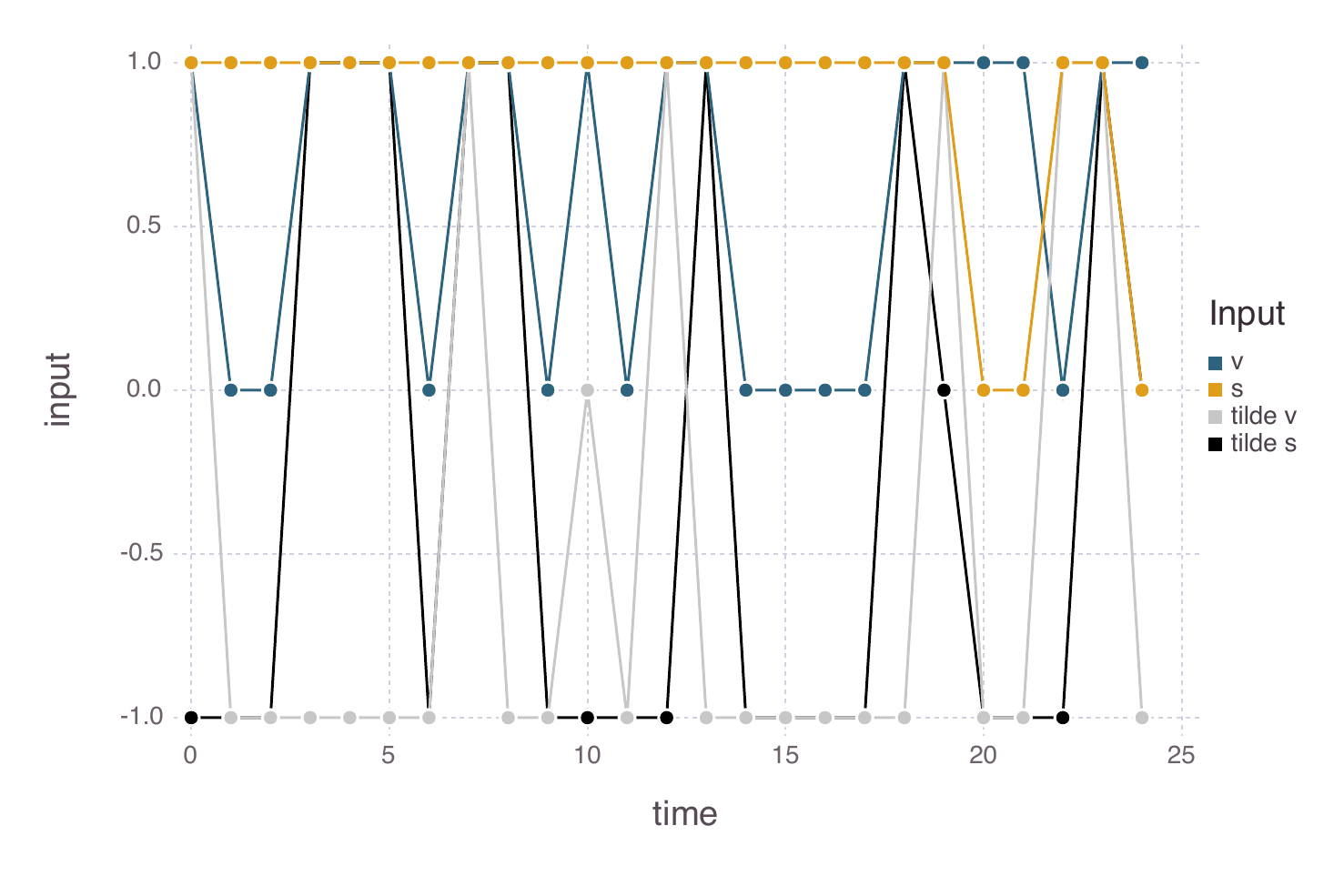}
\end{center}
\caption{Optimal input sequence for cooperative stochastic Dubins vehicle model with $\sigma = 0.3$.}
\label{fig:Dubins_cooperative_input} 
\end{figure}

\begin{figure}[ht!]
\begin{center}
\includegraphics[width=\columnwidth]{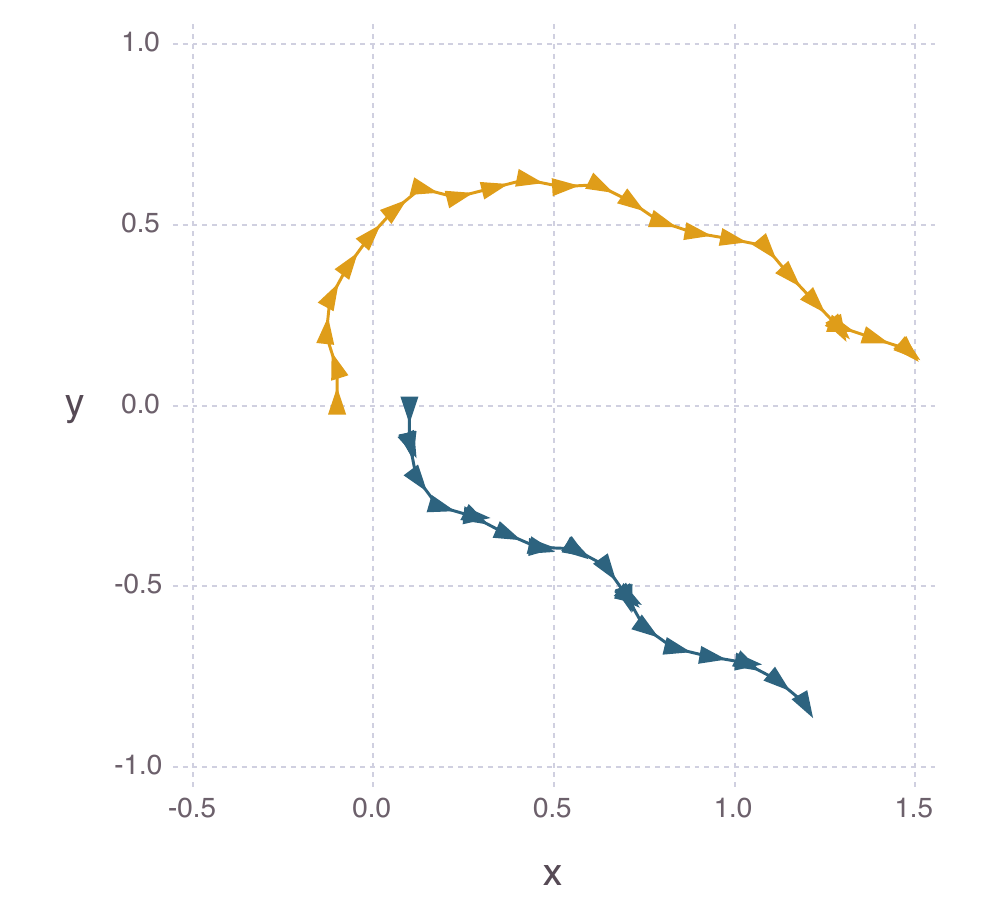}
\end{center}
\caption{Optimal state sequence for the cooperative stochastic Dubins vehicle model with $\sigma = 0.3$. 
}
\label{fig:Dubins_cooperative_state} 
\end{figure}

\begin{figure}[ht!]
\begin{center}
\includegraphics[width=\columnwidth]{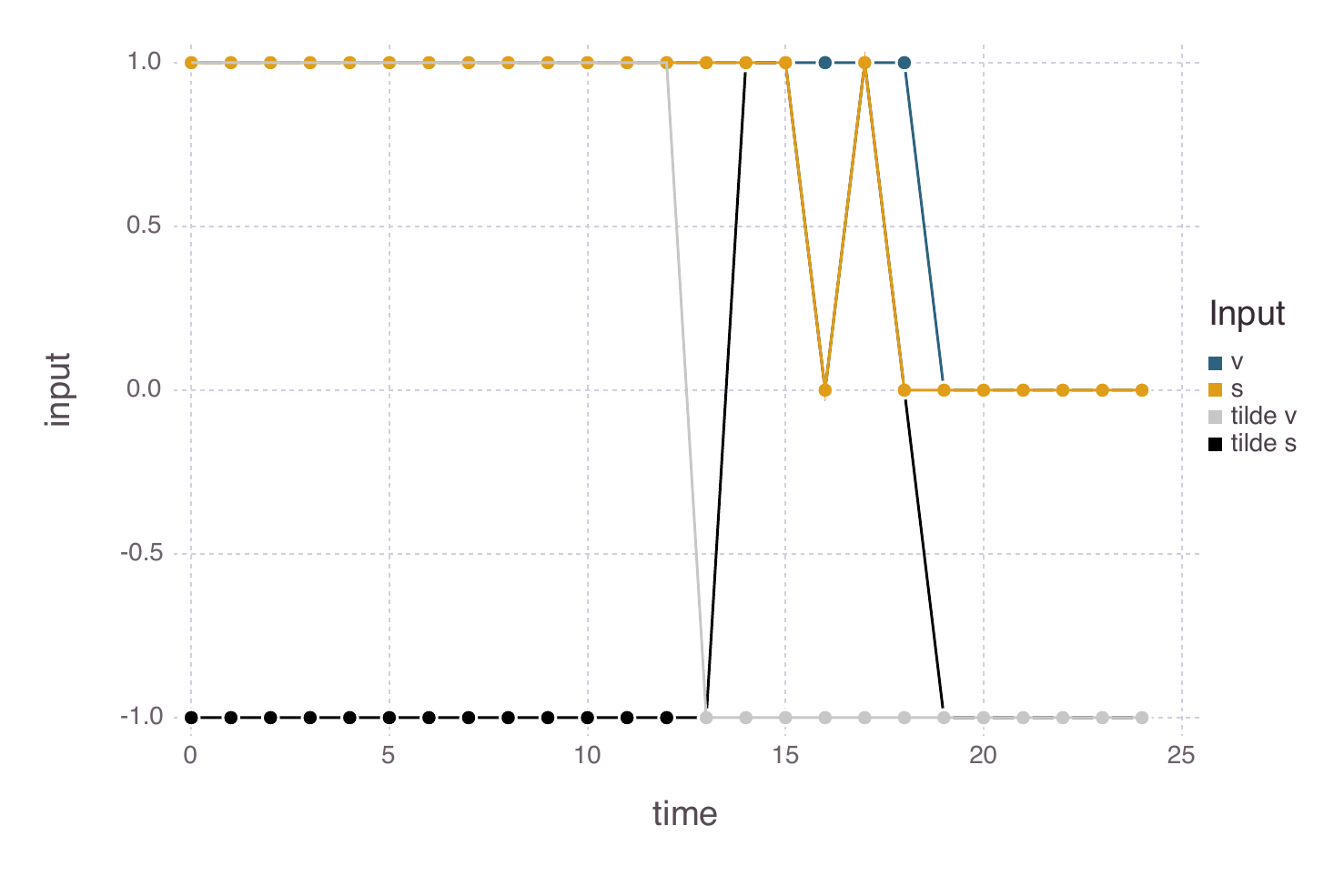}
\end{center}
\caption{Optimal input sequence for cooperative deterministic Dubins vehicle model.}
\label{fig:Dubins_cooperative_input_deterministic} 
\end{figure}

\begin{figure}[ht!]
\begin{center}
\includegraphics[width=\columnwidth]{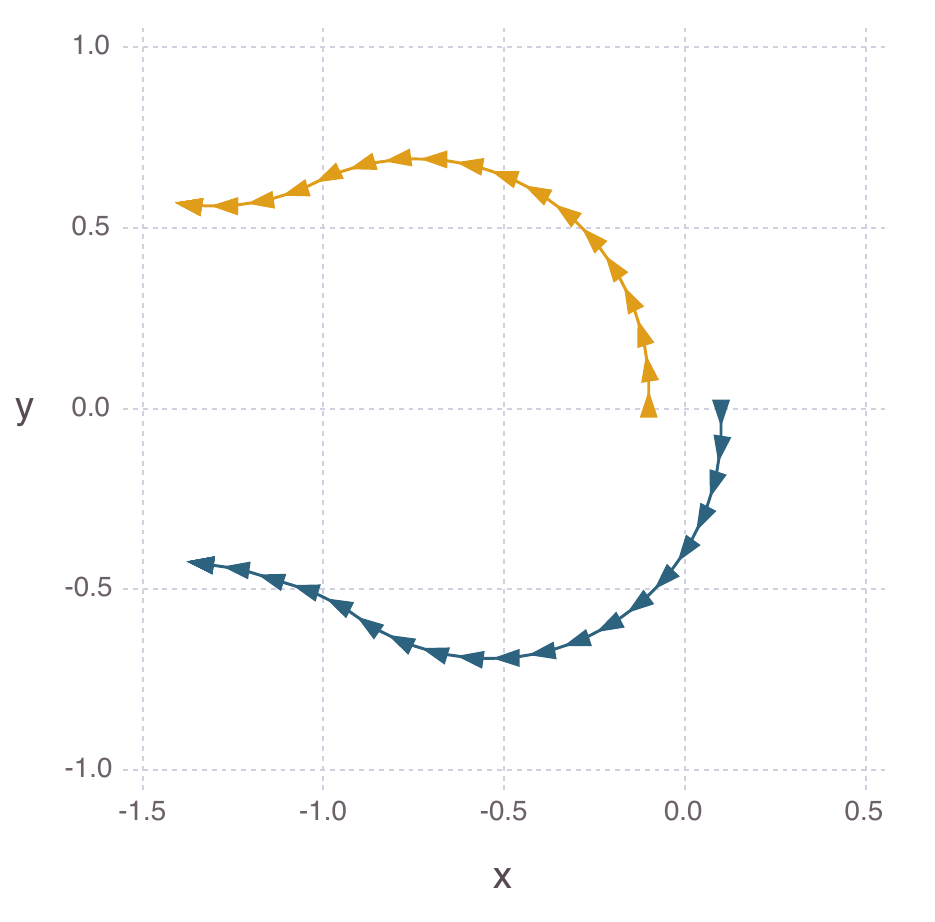}
\end{center}
\caption{Optimal state sequence for the cooperative deterministic Dubins vehicle model. 
}
\label{fig:Dubins_cooperative_state_deterministic} 
\end{figure}

For numerical simulations, the cost functions $\bar J_k$ were computed on a fixed grid of dimension $51 \times 51 \times 65$ using turning radius parameter $L = 1$, input sets $v_k \in \{-0.1, 0, -0.1\}$ and $s_k \in \{-1, 0, -1\}$ Globally optimal input and state trajectory sequences corresponding to the initial condition $x_0 = \begin{bmatrix}0.1 & 0 & \frac{1}{2} \pi & -0.1 & 0 & \frac{3}{2}\pi\end{bmatrix}^T$ are shown in Figures \ref{fig:Dubins_cooperative_input} and \ref{fig:Dubins_cooperative_state}. These are compared against a deterministic version of the model with $w_k = 0$ in Figures  \ref{fig:Dubins_cooperative_input_deterministic} and \ref{fig:Dubins_cooperative_state_deterministic}. 

\section{Application II: MRI Fingerprinting} 
\label{sec:MRI}

Magnetic resonance imaging (MRI) has traditionally focused on acquisition sequences that are static, in the sense that sequences typically wait for magnetization to return to equilibrium between acquisitions. Recently, researchers have demonstrated promising results based on dynamic acquisition sequences, in which spins are continuously excited by a sequence of random input pulses, without allowing the system to return to equilibrium between pulses. Model parameters corresponding to $T_1$ and $T_2$ relaxation, off-resonance and spin density are then estimated from the sequence of acquired data. This technique, termed magnetic resonance fingerprinting (MRF), has been shown to increase the sensitivity, specificity and speed of magnetic resonance studies \citep{Ma13, Davies14}. 

This technique could be further improved by replacing randomized input pulse sequences with sequences that have been optimized for informativeness about model parameters. To this end, we present a model of MR spin dynamics that describes the measured data as a function of $T_1$ and $T_2$ relaxation rates and the sequence of radio-frequency (RF) input pulses, used to excite the spins. 

The following model was introduced in the conference paper \citep{Maidens17-ACC}. In this paper, an optimal control was computed via dynamic programming on a very sparse six-dimensional grid. Now using our symmetry reduction technique, we exploit symmetry reduction to provide a much more accurate optimal input sequence computed on a finer five-dimensional grid. 

We model the spin dynamics via the equations 
\begin{equation}
\mathbf{x}_{k+1} = U_k \begin{bmatrix} \theta_2 & 0 & 0 \\ 0 & \theta_2 & 0 \\ 0 & 0 & \theta_1 \end{bmatrix} \mathbf{x}_k + \begin{bmatrix} 0 \\ 0 \\ 1 - \theta_1 \end{bmatrix}
\label{eq:MRI} 
\end{equation} 
where the states $x_{1, k}$ and $x_{2, k}$ describe the transverse magnetization (orthogonal to the applied magnetic field) and $x_{3, k}$ describes the longitudinal magnetization (parallel to the applied magnetic field). To simplify the presentation, off-resonance is neglected in this model. Control inputs $U_k \in$ SO(3) describe flip angles corresponding to RF excitation pulses that rotate the state about the origin. Between acquisitions, transverse magnetization decays according to the parameter $\theta_2 = e^{- \Delta t/T_2}$ and the longitudinal magnetization recovers to equilibrium (normalized such that the equilibrium is $x_0 = [0 \ \ 0  \ \ 1]^T$) according to the parameter $\theta_1 = e^{- \Delta t/T_1}$ where $\Delta t$ is the sampling interval. 
\begin{figure*}[ht!]
\normalsize
\begin{equation}
\begin{split}
f_k(x_k, U_k, w_k) &=
\begin{bmatrix} 
         &         &         &    0    &    0    &    0     \\
         &   U_k   &         &    0    &    0    &    0    \\ 
         &         &         &    0    &    0    &    0    \\ 
    0    &    0    &    0    &         &         &         \\ 
    0    &    0    &    0    &         &   U_k   &          \\ 
    0    &    0    &    0    &         &         &      
\end{bmatrix}
\begin{bmatrix} 
\theta_2 &    0    &    0    &    0    &    0    &    0      \\
    0    &\theta_2 &    0    &    0    &    0    &    0      \\ 
    0    &    0    &\theta_1 &    0    &    0    &    0      \\ 
    0    &    0    &    0    &\theta_2 &    0    &    0      \\ 
    0    &    0    &    0    &    0    &\theta_2 &    0      \\ 
    0    &    0    &    1    &    0    &    0    &\theta_1  
\end{bmatrix}
x_k
+ 
\begin{bmatrix} 0 \\ 0 \\ 1-\theta_1 \\ 0 \\ 0 \\ -1 \end{bmatrix} \\
g_k(x_k, U_k, w_k) &= 
-x_k^T \begin{bmatrix} 
    0    &    0    &    0    &    0    &    0    &    0   \\
    0    &    0    &    0    &    0    &    0    &    0   \\ 
    0    &    0    &    0    &    0    &    0    &    0   \\ 
    0    &    0    &    0    &\frac{1}{\gamma}&    0    &    0    \\ 
    0    &    0    &    0    &    0    &\frac{1}{\gamma}&    0    \\ 
    0    &    0    &    0    &    0    &    0    &    0    
\end{bmatrix}
x_k
\end{split}
\label{eq:f_and_g}
\end{equation}
\hrulefill
\vspace*{4pt}
\end{figure*}

We assume that data are acquired immediately following the RF pulse, allowing us to make a noisy measurement of the transverse magnetization. We also assume that the measured data are described by a multivariate Gaussian random variable 
\[
\mathbf y_k = \begin{bmatrix} 1 & 0 & 0 \\ 0 & 1 & 0 \end{bmatrix}\mathbf x_k  + v_k 
\]
where $v_k$ is a zero-mean Gaussian noise with covariance $\begin{bmatrix} \gamma & 0 \\ 0 & \gamma \end{bmatrix}$. This model results from a time discretization of the Bloch equations \citep{Bloch46, Nishimura10} under a time scale separation assumption that specifies that the RF excitation pulses act on a much faster time scale than the relaxation time constants $T_1$ and $T_2$. A simplified two-state version of this model was considered in \citep{Maidens16-CDC}, where the transverse magnetization was modelled using a single state describing the magnitude of $[x_{1, k}, x_{2, k}]^T$. 

We see from the model \eqref{eq:MRI} that magnetization in the transverse direction decays while magnetization in the longitudinal direction grows. However only the transverse component of the magnetization can be measured. Thus there is a trade-off between making measurements (which leads to loss of magnetization) and magnetization recovery. This is the trade-off that we hope to manage through the optimal design of an input sequence $U_k$. 

We wish to quantify the informativeness of an acquisition sequence based on the information about the $T_1$ relaxation parameter $\theta_1$ that is contained in the resulting data set. More formally, we wish to choose $U_k \in $ SO(3) to maximize the Fisher information about $\theta_1$ contained in the joint distribution of $Y = (\mathbf y_0, \dots, \mathbf y_N)$. The Fisher information $\mathcal{I}$ can be expressed as a quadratic function of the sensitivities of $\mathbf x_k$ with respect to $\theta_1$:
\[
\mathcal I = \sum_{k=0}^N 
 \frac{\partial}{\partial \theta_1} \mathbf{x}_k^T
\begin{bmatrix} 1/\gamma & 0 & 0 \\ 0 & 1/\gamma & 0 \\ 0 & 0 & 0 \end{bmatrix} 
\frac{\partial}{\partial \theta_1} \mathbf{x}_k
\]
where the sensitivities $ \frac{\partial}{\partial \theta_1} \mathbf{x}_k$ satisfy the following sensitivity equations: 
\[
\resizebox{\hsize}{!}{$
\frac{\partial}{\partial \theta_1} \mathbf{x}_{k+1} = U_k \begin{bmatrix} \theta_2 & 0 & 0 \\ 0 & \theta_2 & 0 \\ 0 & 0 & \theta_1 \end{bmatrix} \frac{\partial}{\partial \theta_1} \mathbf{x}_k + U_k \begin{bmatrix} 0 & 0 & 0 \\ 0 & 0 & 0 \\ 0 & 0 & 1 \end{bmatrix} \mathbf{x}_k + \begin{bmatrix}0 \\ 0 \\ -1\end{bmatrix}. 
$}
\]

It should be noted that for system \eqref{eq:MRI}, the objective function $\mathcal{I}$ has many local optima as a function of the input sequence $U_k$. Thus, in contrast with \citep{Maidens16-TMI} which consider optimal experiment design for hyperpolarized MRI problems, for this model, local search methods provide little insight into what acquisition sequences are good. In contrast with the MRI model presented in \citep{Maidens16-ACC}, where global optimal experiment design heuristics are developed for linear dynamical systems, in this model the decision variables $U_k$ multiply the state vector $\mathbf x_k$, making the output $\mathbf y_k$ a nonlinear function of the sequence $U = (U_0, \dots U_{k-1})$. Thus we must use dynamic programming to find a solution.

\subsection{Model} 

To present this problem in the formalism we have introduced, we define an augmented state vector 
\[
x_k = \begin{bmatrix} \mathbf{x}_k \\ \frac{\partial}{\partial \theta_1} \mathbf{x}_k \end{bmatrix} \in \mathbb{R}^6.
\]
We can write the dynamics of the augmented state as a control system with $f$ and $g$ defined in Equation \eqref{eq:f_and_g}.  
This system has a one-dimensional group of symmetries defined by 
\[
\resizebox{\hsize}{!}{$
\begin{split}
\phi_\alpha(x_k) &= 
\begin{bmatrix} 
\cos(\alpha)&-\sin(\alpha)&    0    &    0    &    0    &    0    \\
\sin(\alpha)& \cos(\alpha)&    0    &    0    &    0    &    0     \\ 
    0    &    0    &    1    &    0    &    0    &    0     \\ 
    0    &    0    &    0    &\cos(\alpha)&-\sin(\alpha)&    0     \\ 
    0    &    0    &    0    &\sin(\alpha)&\cos(\alpha)&     0     \\ 
    0    &    0    &    0    &    0    &    0    &    1       
\end{bmatrix} x_k \\
\chi_\alpha(U_k) &= \begin{bmatrix} \cos(\alpha) & -\sin(\alpha) & 0 \\
                             \sin(\alpha) &  \cos(\alpha) & 0 \\
                             0 & 0 & 1 
             \end{bmatrix} 
             U_k
             \begin{bmatrix} \cos(\alpha) & \sin(\alpha) & 0 \\
                             -\sin(\alpha) &  \cos(\alpha) & 0 \\
                             0 & 0 & 1 
             \end{bmatrix}\\
\psi_\alpha(w_k) &= w_k
\end{split}
$}
\]
for any $\alpha \in \mathbb{R}/2\pi\mathbb{Z}$. 

\subsection{Dynamic programming in reduced coordinates} 
\label{sec:reduced}

To perform dynamic programming in a reduced coordinate system, we begin by defining the cross-section $\mathcal{C} = \{x: x_1 = 0, x_2 > 0\}$, and computing the moving frame $\gamma(x)$. To do so, we solve 
\[
0 = \phi^a_{\gamma(x)} (x) = x_1 \cos \gamma(x) - x_2 \sin \gamma(x). 
\]
Isolating $\gamma$ yields 
\[
\gamma(x) = \operatorname{atan2}(x_1, x_2)
\]
where $\operatorname {atan2}$ denotes the multi-valued inverse tangent function
\[
\operatorname {atan2} (y,x)={\begin{cases}\arctan({\frac {y}{x}})&{\text{if }}x>0,\\\arctan({\frac {y}{x}})+\pi &{\text{if }}x<0{\text{ and }}y\geq 0,\\\arctan({\frac {y}{x}})-\pi &{\text{if }}x<0{\text{ and }}y<0,\\+{\frac {\pi }{2}}&{\text{if }}x=0{\text{ and }}y>0,\\-{\frac {\pi }{2}}&{\text{if }}x=0{\text{ and }}y<0,\\{\text{undefined}}&{\text{if }}x=0{\text{ and }}y=0.\end{cases}}
\]

Next, we compute the invariants $\rho(x)$ using
\begin{equation*}
\resizebox{\columnwidth}{!}{$
\begin{split}
\rho(x) &= \phi^b_{\gamma(x)} \\
&= 
\begin{bmatrix} 
\sin(\operatorname{atan2}(x_1, x_2))& \cos(\operatorname{atan2}(x_1, x_2))&    0    &    0    &    0    &    0     \\ 
    0    &    0    &    1    &    0    &    0    &    0     \\ 
    0    &    0    &    0    &\cos(\operatorname{atan2}(x_1, x_2))&-\sin(\operatorname{atan2}(x_1, x_2))&    0     \\ 
    0    &    0    &    0    &\sin(\operatorname{atan2}(x_1, x_2))&\cos(\operatorname{atan2}(x_1, x_2))&     0     \\ 
    0    &    0    &    0    &    0    &    0    &    1       
\end{bmatrix} x \\
&= \begin{bmatrix} \sqrt{x_1^2 + x_2^2} \\ 
x_3 \\ 
\frac{1}{\sqrt{x_1^2 + x_2^2}} (x_2 x_4 - x_1 x_5) \\
\frac{1}{\sqrt{x_1^2 + x_2^2}} (x_1 x_4 + x_2 x_5) \\
x_6 \end{bmatrix}. 
\end{split}
$}
\end{equation*}

Further, $\rho$ restricted to the cross-section $\mathcal{C}$ is injective with inverse $\bar \rho^{-1}: \mathbb{R}_+ \times \mathbb{R}^4 \to \mathcal{C}$ given by 
$
\bar \rho^{-1}(\bar x) = \begin{bmatrix} 0 & \bar x_1 & \bar x_2 & \bar x_3 & \bar x_4 & \bar x_5 \end{bmatrix}^T. 
$ The theory above tells us we can thus solve the optimal stochastic control problem in a 5 dimensional state space, reducing the original 6 dimensional problem of 1 dimension. 

Note that this reduction can be computed using only the state transformation group $\phi$, without reference to the state update equation $f$. Thus the same reduction can be applied to any system with the same symmetries. 

\subsection{Results} 

To implement this algorithm, we discretize the reduced five-dimensional state space and two-dimensional input space via grids of size $6 \times 10 \times 15 \times 15 \times 15$ and $16 \times 8$ respectively. The code was written in the Julia language and parallelized to allow evaluation of $J_k$ in parallel across grid points \citep{Maidens16-CDC}. The implementation is publicly available at \url{https://github.com/maidens/Automatica-2017}. 

\begin{figure}[ht!]
\begin{center}
\includegraphics[width=\columnwidth]{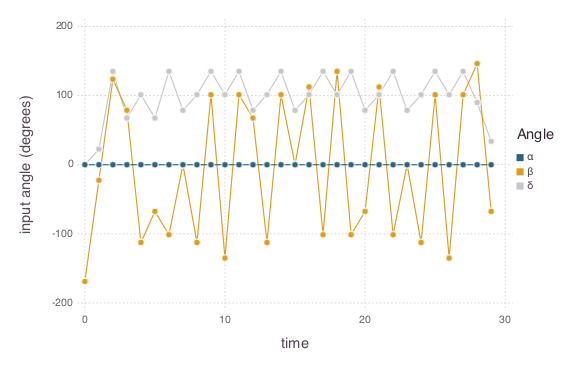}
\end{center}
\caption{Optimal input sequence for the MR fingerprinting model. The angles $\alpha$, $\beta$ and $\delta$ represent rotations about the $z$, $y$ and $x$ axes respectively, resulting in an control input $U_k = R_z(\alpha_k) R_y(\beta_k) R_x(\gamma_k)$.}
\label{fig:MRI_optimal_input} 
\end{figure}

\begin{figure}
    \centering
    \begin{subfigure}[b]{\columnwidth}
        \includegraphics[width=\textwidth]{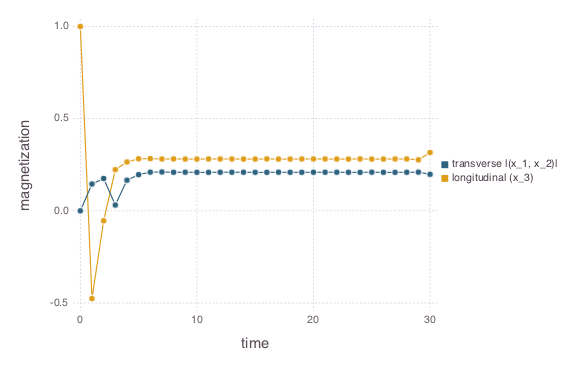}
        \caption{Magnetizations}
        \label{fig:MRI_optimal_magnetization}
    \end{subfigure}

    \begin{subfigure}[b]{\columnwidth}
        \includegraphics[width=\textwidth]{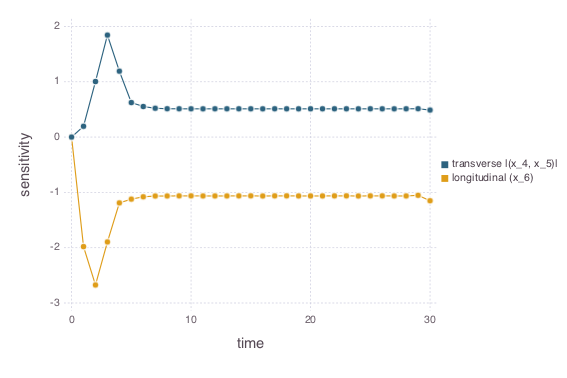}
        \caption{Sensitivities}
        \label{fig:MRI_optimal_sensitivity}
    \end{subfigure}
    \caption{Optimal state sequence for the MR fingerprinting model. Here we have plotted the longitudinal and transverse components of both the magnetization (states $x_1$, $x_2$, and $x_3$) and the sensitivities (states $x_4$, $x_5$, and $x_6$) where the transverse component is computed as the Euclidean norm of the vectors $(x_1, x_2)$ and $(x_4, x_5)$ respectively. }\label{fig:MRI_optimal_states}
\end{figure}

Optimal input and state trajectories for the model corresponding to the initial condition at the equilibrium $x_0 = [0 \ \  0 \ \ 1\ \  0\ \   0\ \   0]^T$ are plotted in Figures \ref{fig:MRI_optimal_input} and \ref{fig:MRI_optimal_states}.

In contrast with the results from \citep{Maidens16-CDC} where we considered a simplified version of the model, for this full model we no longer find that the optimal flip angle sequence converges to a cyclic pattern, rather it appears irregular. However, state sequence of longitudinal magnetizations and transverse magnetization magnitudes appears to converge to a constant sequence. This is likely because in this work we assumed Gaussian noise in the inputs in contrast with the Rician noise assumed in the previous work, therefore it is no longer necessary to conserve magnetization across multiple time steps before generating a reliable measurement.

\section{Conclusion} 
\label{sec:conclusion} 

We have presented a method of reducing the complexity of dynamic programming for systems in which the state dynamics, stage costs and transition probabilities are invariant under a group of symmetries. This allows us to compute globally optimal control policies for systems of moderate state dimension. We have applied this technique to compute globally optimal trajectories to a six-dimensional original MRI model with a one-dimensional group of symmetries and for a six-dimensional stochastic Dubins vehicle model with a three-dimensional group of symmetries  by reducing the dimension of the state space to five and three dimensions respectively. Since computation time for dynamic programming depends exponentially on the state space dimension, this technique enables the computation of optimal control policies for systems in which it was previously infeasible. 

\begin{ack}                           
Research supported in part by the National Science Foundation under grant ECCS-1405413. 
\end{ack}

\bibliographystyle{abbrvnat}
\bibliography{references}         

\end{document}